\documentclass[a4paper,12pt,fleqn]{scrartcl}
\usepackage{graphicx}
\usepackage{amsmath}
\usepackage{amssymb}
\usepackage{bm}
\usepackage{fourier}
\usepackage{hyperref}
\setkomafont{sectioning}{\fontencoding{OT1}\fontfamily{phv}\fontshape{sf}\fontseries{b}\selectfont}
\hypersetup{colorlinks=true,urlcolor=blue}

\begin{document}

\section*{Bringing order to the expanding fermion zoo\footnote{Perspective published in: Science \textbf{353}, 539--540 (2016) \href{http://dx.doi.org/10.1126/science.aag2865}{DOI:10.1126/science.aag2865}}}
{\large Crystallography provides an inventory of the electron-like particles\smallskip\\ 
that emerge in a lattice world.}\bigskip\\
\textit{Carlo Beenakker, Leiden University}\medskip

We observe space as a continuum, but we might entertain the thought that there is an underlying lattice and that space is actually a crystal. Which particles would inhabit such a lattice world? This question was first raised by Werner Heisenberg in 1930, in an attempt to remove the infinities that plagued the continuum quantum mechanics. His \textit{Gitter\-welt} (lattice world) hosted electrons that could morph into protons, photons that were not massless, and more peculiarities that compelled him to abandon ``this completely crazy idea'' \cite{Hei30}. Heisenberg's motivation to put electrons on a lattice came from solid state physics, which in the 1930's was just developing as a field of research, and which has now become a playground for ``crazy ideas'' from particle physics. In this spirit Barry Bradlyn and colleagues \cite{Bra16} have used crystallography to classify the electronic excitations of the lattice world, identifying materials where they become a reality as (nonfundamental) quasiparticles.

The idea that electrons on a lattice might turn into altogether different quasiparticles was forcefully demonstrated in graphene: An electron moving on the two-dimensional honeycomb lattice of carbon atoms looses its mass. In the language of particle physics, the electron is transformed from a massive Dirac fermion into a massless Weyl fermion. If the lattice is superconducting the electron may loose its charge, becoming a neutral Majorana fermion. These three types of fermions exhaust the options for a spin-$1/2$ particle in continuous space, but lattices offer more possibilities. Recent additions \cite{Bur11,Hei15,Cha16,Wan16,Wie16,Sol15,Mue16} to the fermion family go by the names of nodal-line fermions, nexus fermions, hourglass fermions, double-Dirac fermions, tilted fermions --- each stabilized by a different lattice symmetry and characterized by a different energy-momentum relationship.

To bring order into this zoo of lattice fermions Bradlyn \textit{et al.} turned to the symmetry classification of crystals: A table that groups all known crystal structures (more than half a million) into one of 230 ``space groups'', depending on their translational, rotational, and reflection symmetry. The proof that there are no more and no less than 230 distinct ways to combine these symmetries in three-dimensional space (in two dimensions there are only 17 ways) was a \textit{tour-de-force} of nineteenth century crystallography. Weyl fermions appear in any space group without inversion symmetry, but only a few space groups (a total of 16 to be precise) have the right combination of symmetries to stabilize the lattice fermions.

\begin{figure}[tb]
\centerline{\includegraphics[width=0.5\linewidth]{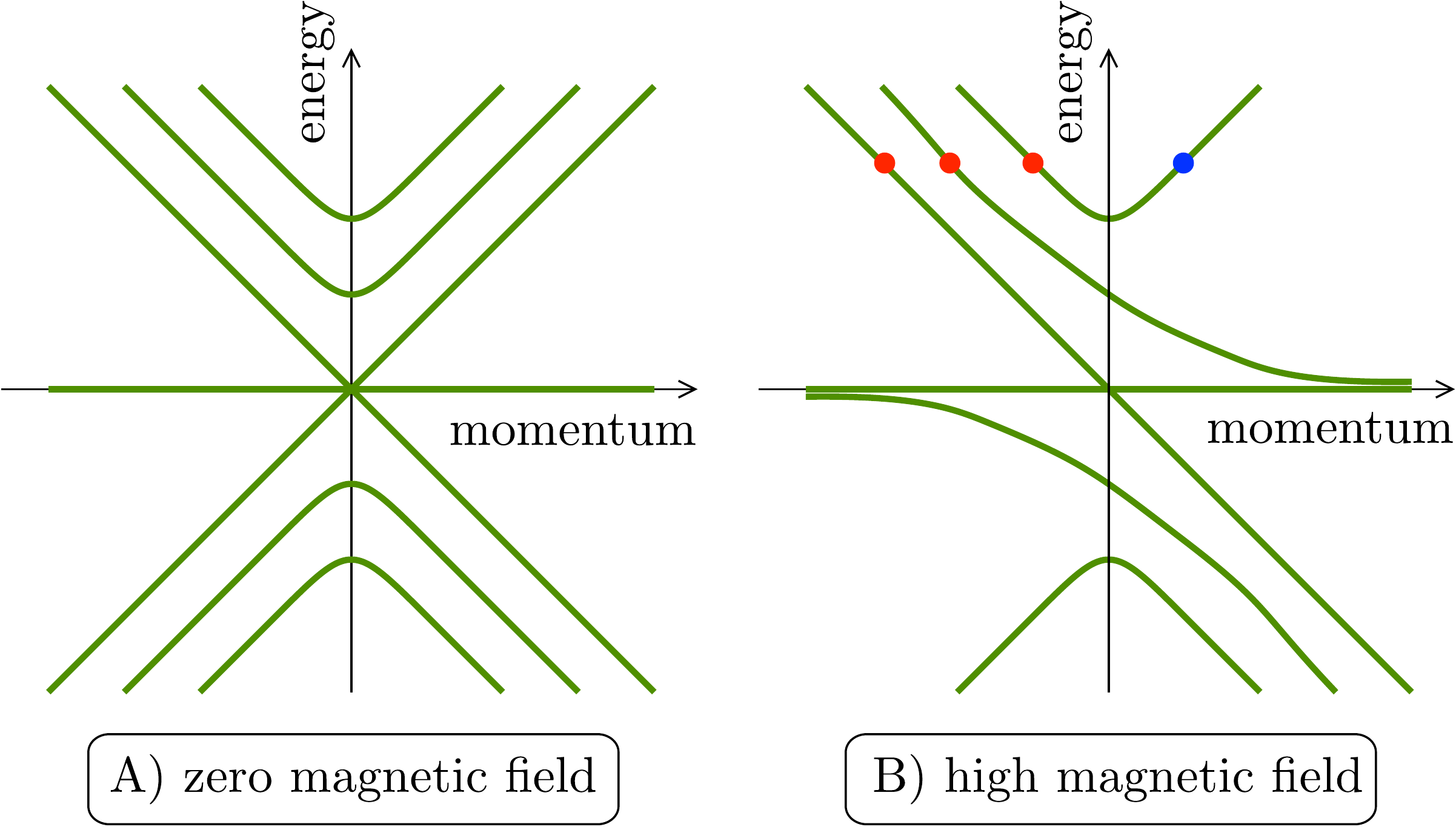}}
\caption{Energy-momentum relation of a spin-1 Weyl fermion, in zero magnetic field (panel A) and in a high field parallel to the momentum (panel B). The flat band at zero energy appears when the lattice augments the spin from $1/2$ to $1$. The slope of the curves gives the direction of motion. At a given energy in panel B there are $n_{\rm L}=3$ left-movers (red dots) and $n_{\rm R}=1$ right-movers (blue), indicating a Chern number $C=n_{\rm L}-n_{\rm R}=2$. For spin-$1/2$ Weyl fermions, in contrast, $C=1$.
}
\end{figure}

A symmetry operation that is particularly effective at preventing the opening of an excitation gap \cite{You15} is the combination of a translation by a fraction of the unit cell with a rotation or reflection, resulting in a screw-rotation or a glide-reflection symmetry (the technical term is ``nonsymmorphic'' symmetry). Bradlyn \textit{et al.} show that this can transform the spin-$1/2$ Weyl fermion into a spin-1 quasiparticle. It remains a fermion --- the requirement that particles with integer spin are bosons applies only to fundamental particles. While a spin-1/2 particle is represented by a two-component wave function (its magnetic moment can be $-1/2$ or $+1/2$), for spin-1 we need three components (magnetic moment $-1,0,+1$). Two components produce a linear crossing of energy bands, the third component intersects it with a nearly flat band (see the figure). Such a three-band crossing point was observed in a zinc-blende crystal \cite{Orl14}, accompanied by a massless excitation named ``Kane fermion'', but there fine-tuning to a critical point of the phase diagram was required to avoid a gap opening. The gapless spectrum of a spin-1 Weyl fermion is enforced by lattice symmetry, without any fine-tuning of parameters. 

Because the lattice symmetries are broken when the crystal terminates, surfaces can locally destabilize the lattice fermions. Topology comes to the rescue, protecting some of the quasiparticles with a conserved quantity called the Chern number $C$. In a magnetic field $C$ counts the difference between the number of left-moving and right-moving states, relative to the magnetic field direction. The excess number of left- or right-movers cannot terminate at a boundary, it must continue as a gapless surface state called a ``Fermi arc''. The spin-1/2 Weyl fermions have $C= 1$, producing a single Fermi arc, the new spin-1 variety has $C= 2$ with two Fermi arcs. An unusual magnetoresistance is expected from the imbalance of left-movers and right-movers, by analogy with the ``chiral anomaly'' of Weyl fermions \cite{Nie83}.

There is no shortage of crystals in which these phenomena might be observed, at least based on numerical calculations of the band structure. Bradlyn \textit{et al.} suggest half a dozen hosts for the spin-1 Weyl fermion, and experiments to detect it with spectroscopy are underway. For detection by transport experiments one would like to have the Fermi level at or close to the band crossing point (the Weyl point). In graphene this alignment is ensured by charge neutrality, but that is a special property of the two-dimensional carbon lattice which does not carry over to three dimensions. Fortunately, the band structure calculations indicate that in the most promising materials the Fermi level is only a few tenths of an electron volt from the Weyl point, so that signatures of the singularity should be readily observable in electrical conduction.

Stepping back from the lattice world, one might ask: why not discretize time as well as space and put space-time on a lattice? A reflection in the temporal direction corresponds to time reversal, with this additional symmetry operation the 230 space groups expand to a total of 1651 socalled magnetic space groups. The fermion zoo may expand quite a bit further.

{\small
\noindent {\sc\bfseries acknowledgments} ---
I thank Charles Kane and Titus Neupert for insightful comments and an ERC Synergy grant for support.
}


\begin{thebibliography}{99}
\bibitem{Hei30} W. Heisenberg, correspondence cited and discussed in B. Carazza and H. Kragh, \textit{Heisenberg's lattice world: The 1930 theory sketch}, Am. J. Phys. \textbf{63}, 595 (1995).
\bibitem{Bra16} B. Bradlyn, J. Cano, Z. Wang, M. G. Vergniory, C. Felser, R. J. Cava, and B. A. Bernevig, \textit{Beyond Dirac and Weyl fermions: Unconventional quasiparticles in conventional crystals}, Science \textbf{353} (\href{http://dx.doi.org/10.1126/science.aaf5037}{in press}, 2016).
\bibitem{Bur11} A. A. Burkov, M. D. Hook, and L. Balents, \textit{Topological nodal semimetals}, Phys. Rev. B \textbf{84}, 235126 (2011).
\bibitem{Hei15} T. T. Heikkil\"{a} and G. E. Volovik, \textit{Nexus and Dirac lines in topological materials}, New J. Phys. \textbf{17}, 093019 (2015).
\bibitem{Cha16} Guoqing Chang, Su-Yang Xu, Shin-Ming Huang, Daniel S. Sanchez, Chuang-Han Hsu, Guang Bian, Zhi-Ming Yu, Ilya Belopolski, Nasser Alidoust, Hao Zheng, Tay-Rong Chang, Horng-Tay Jeng, Shengyuan A. Yang, Titus Neupert, Hsin Lin, and M. Zahid Hasan, \textit{New fermions on the line in topological symmorphic metals}, arXiv:1605.06831.
\bibitem{Wan16} Zhijun Wang, A. Alexandradinata, R. J. Cava, and B. A. Bernevig, \textit{Hourglass fermions}, Nature \textbf{532}, 189 (2016).
\bibitem{Wie16} B. J. Wieder, Y. Kim, A. M. Rappe, and C. L. Kane, \textit{Double Dirac semimetals in three dimensions}, Phys. Rev. Lett. \textbf{116}, 186402 (2016).
\bibitem{Sol15} A. A. Soluyanov, D. Gresch, Zhijun Wang, QuanSheng Wu, M. Troyer, Xi Dai, B. A. Bernevig, \textit{Type-II Weyl semimetals}, Nature \textbf{527}, 495-498 (2015).
\bibitem{Mue16} L. Muechler, A. Alexandradinata, T. Neupert, and R. Car, \textit{Tilted Dirac fermions}, arXiv:1604.01398.
\bibitem{You15} S. M. Young and C. L. Kane, \textit{Dirac semimetals in two dimensions}, Phys. Rev. Lett. \textbf{115}, 126803 (2015).
\bibitem{Orl14} M. Orlita, D. M. Basko, M. S. Zholudev, F. Teppe, W. Knap, V. I. Gavrilenko, N. N. Mikhailov, S. A. Dvoretskii, P. Neugebauer, C. Faugeras, A-L. Barra, G. Martinez, and M. Potemski, \textit{Observation of three-dimensional massless Kane fermions in a zinc-blende crystal}, Nature Phys. \textbf{10}, 233 (2014).
\bibitem{Nie83} H. B. Nielsen and M. Ninomiya, \textit{The Adler-Bell-Jackiw anomaly and Weyl fermions in a crystal}, Phys. Lett. B \textbf{130}, 389 (1983).
\end{thebibliography}
\end{document}